\documentclass[showpacs,twocolumn,amsmath,amssymb,pre,aps]{revtex4} 
\usepackage{graphicx}
\usepackage{bm}
\usepackage{mathrsfs}

\begin{document}
\title{
Enhancing the critical current in quasiperiodic pinning arrays \\ 
below and above the matching magnetic flux 
}
\draft

\author{V.R.~Misko$^{1,2}$,
D.~Bothner$^{3}$, M.~Kemmler$^{3}$, R.~Kleiner$^{3}$, 
D.~Koelle$^{3}$, F.M.~Peeters$^{1}$, and Franco Nori$^{2,4}$}
\affiliation{$^{1}$Departement Fysica, Universiteit Antwerpen, 
B-2020 Antwerpen, Belgium}
\affiliation{$^2$ Advanced Science Institute, RIKEN, Wako-shi, Saitama 351-0198, Japan}
\affiliation{$^{3}$Physikalisches Institut-Experimentalphysik II 
and Center for Collective Quantum Phenomena, 
Universit\"{a}t T\"{u}bingen, D-72076 T\"{u}bingen, Germany}
\affiliation{$^4$ Physics Department, University of Michigan, 
Ann Arbor, MI 48109-1040, USA}

\date{\today}

\begin{abstract}
Quasiperiodic pinning arrays, as recently demonstrated theoretically
and experimentally using a five-fold Penrose tiling, can lead to a significant 
enhancement of the critical current $I_{c}$  as compared to ``traditional'' 
regular pinning arrays.
However, while regular arrays showed only a sharp peak in 
$I_{c}(\Phi)$
at the matching flux $\Phi_{1}$ and quasiperiodic arrays provided
a much broader maximum at $\Phi < \Phi_{1}$, both types of pinning
arrays turned out to be inefficient for fluxes larger than
$\Phi_{1}$.
We demonstrate theoretically and experimentally the enhancement
of $I_{c}(\Phi)$ for $\Phi > \Phi_{1}$ by using non-Penrose
quasiperiodic pinning arrays.
This result is based on a qualitatively different mechanism 
of flux pinning by quasiperiodic pinning arrays 
and could be potentially useful for applications in superconducting 
micro-electronic devices operating in a broad range of magnetic fields.
\end{abstract}
\maketitle

\section{Introduction} 

Pinning of Abrikosov vortices in superconductors is a topic 
that has attracted condsiderable interest, both with respect 
to the fundamental physical properties of so-called 
``vortex-matter'' and with respect to device applications. 
In the latter case, for bulk or thick film superconductors, 
the inclusion of nanodefects, acting as pinning sites, typically 
randomly distributed within the superconductors, has been 
shown to significantly enhance the critical current density, 
which is important for increasing the current-carrying capacity 
of wires and tapes, e.g., for applications in superconducting 
magnets (see, e.g., Ref.~\cite{Gutierrez07} and references therein). 
On the other hand, recent progress in the fabrication of 
nanostructures has provided a wide variety of well-controlled 
lithographically defined artificial pinning sites, and 
experiments can control both the location and the strength 
of each pinning center in vortex-confining Arrays of Pinning 
Sites (APS) (see, e.g., \cite{vvmdotprb,fnsc2003,rwdot,wendr}). 
In this context it has been shown that e.g. microholes 
(antidots) or magnetic dots may improve the performance 
of micro-electronic thin film devices. 
Examples are the suppression of low-frequency flux noise 
in superconducting quantum interference devices (SQUIDs) 
by strategically positioned antidots 
\cite{Roger2000,Roger2002}
or the enhancement of the quality factor 
of superconducting microwave resonators operated in magnetic 
fields, which can be achieved by incorporating antidot arrays 
into the thin film structures \cite{Bushev}. 
From a basic point of view, it is interesting to explore 
the optimum pinning landscape provided by APS. 
One convenient way to do this is to investigate the critical 
current $I_{c}$ versus magnetic flux $\Phi$ for different 
arrangements of pinning sites. 

Periodic artificial APS were first proposed as devices
able of trapping magnetic flux and thus enhancing $I_{c}$.
However, the efficiency of periodic APS is strongly limited
to so-called matching fields, i.e., when the number 
of vortices is commensurate with the number of pins, 
thus resulting in strong but very {\it narrow}
peaks in the critical current versus magnetic flux,
$I_{c}(\Phi)$.
This shortcoming can be overcome by using more sophisticated
pinning topologies, e.g., quasiperiodic (QP) APS
\cite{weprl,weprb,koelle,silhanek,villegas,vvmpen2009}. 
In particular, using a five-fold Penrose tiling 
(see, e.g., 
Refs.~\cite{penrose,steinhardtprl84,steinhardtprb97,steinhardtnature,steinhardtbookqc,bookqc,tilings}) 
as an APS
(i.e., placing pins on the vertices of a Penrose tiling) 
results in a considerable enhancement of $I_{c}(\Phi)$,
i.e., significant broadening of the peak in $I_{c}(\Phi)$,
as was recently demonstrated theoretically \cite{weprl,weprb}
and experimentally \cite{koelle,silhanek}.
The underlying idea of using QP APS is that, contrary to
regular APS, they include {\it many} periodicities and
thus can trap vortices for magnetic fluxes other than
matching values $\Phi_{i}$.
As shown for Penrose-tiling APS
\cite{weprl,weprb,koelle,silhanek},
in addition to the matching-flux peak at $\Phi = \Phi_{1}$,
$I_{c}(\Phi)$ acquires a remarkably {\it broad} maximum for
$\Phi < \Phi_{1}$.
This maximum turned out to be even more robust with respect
to variations of parameters of APS than the sharp peak at
$\Phi = \Phi_{1}$ \cite{weprl,weprb}.
However, in theory and experiments, $I_{c}(\Phi)$ decreases
rapidly for fluxes
$\Phi > \Phi_{1}$ for QP Penrose-tiling APS as well as
for regular APS.
Here, we discuss the essential difference in
the flux pinning for magnetic fluxes below and above
the matching value $\Phi = \Phi_{1}$ and we demonstrate,
theoretically and experimentally, how the critical current
$I_{c}(\Phi)$ can be improved in the important regime where
$\Phi > \Phi_{1}$, by using novel artificial APS. 

The paper is organized as follows. 
The model and experiment are described in Sec.~II and Sec.~III, 
respectively. 
In Sec.~IV, we analyze the enhancement of the critical current 
below the first matching field in different aperiodic pinning 
arrays and compare the results to those for a five-fold Penrose-tiling 
APS. 
The mechanisms of vortex ``conductivity'' leading to the enhancement 
of the critical current below and above the matching flux are 
discussed in Sec.~V. 
In Sec.~V, we also present our experimental and numerical results 
for QP APS demonstrating the enhancement of the critical current 
for $\Phi > \Phi_{1}$. 
Finally, the conclusions are given in Sec.~VI.

\section{Simulation} 

We model a three-dimensional (3D) slab, infinitely long in the 
$z$-direction, 
by a 2D (in the $xy$-plane) simulation cell with periodic 
boundary conditions. 
Note that following Refs.~\cite{weprl,weprb} we apply periodic 
boundary conditions at the boundaries of the square simulation cell 
while the quasiperiodic array is taken smaller than the cell. 
This is done in order to prevent imposing {\it ad hoc} periodicity 
to the vortex motion in a QP APS: the free-of-pinning region between 
the QP APS and the boundary of the simulation cell serves as a 
reservoir of vortices that mimics the external applied magnetic 
field. 
This reservoir 
(of thickness of the order of few intervortex distances) 
erases the memory, i.e., the information on the 
coordinate of a vortex leaving the cell. 
Each additional vortex enters the QP sample at the current point 
of ``least resistance to entry'', similarly to the way vortices 
enter any other sample when placed in an external magnetic field. 
This approach has been successfully used in numerous simulations 
in the past (see, e.g., \cite{FNprl94,FNsci97,FNprb97,weprl,weprb}).

To study the dynamics of vortex motion, 
we perform simulated annealing simulations by numerically integrating
the overdamped equations of motion
(see, e.g., Refs.~\cite{weprl,weprb} for a description of the method):
\begin{equation}
\eta {\rm \bf v}_{i} \ = \ {\rm \bf f}_{i} \ = \ {\rm \bf f}_{i}^{vv} + {\rm \bf f}_{i}^{vp} + {\rm \bf f}_{i}^{T} + {\rm \bf f}_{i}^{d}.
\end{equation}
Here
${\rm \bf f}_{i}$
is the total force per unit length acting on vortex
$i$,
${\rm \bf f}_{i}^{vv}$
and
${\rm \bf f}_{i}^{vp}$
are the forces due to vortex-vortex and vortex-pin interactions, 
respectively, 
${\rm \bf f}_{i}^{T}$
is the thermal stochastic force,
and
${\rm \bf f}_{i}^{d}$
is the driving force;
${\rm \bf v}_{i}$
is the velocity and
$\eta$ is the viscosity.
All the forces are expressed in units of
$
f_{0} = \Phi_{0}^{2} / 8 \pi^{2} \lambda^{3},
$
where $\Phi_{0} = hc/2e$, and
lengths (fields) are in units of
$\lambda$ ($\Phi_{0}/\lambda^{2}$). 

Following Refs.~\cite{weprl,weprb}, 
we model vortex pinning by short-range parabolic potential wells 
located at positions 
${\rm \bf r}_{k}^{(p)}$. 
The pinning force is 
\begin{equation}
{\rm \bf f}_{i}^{vp} = \sum\limits_{k}^{N_{p}} \left( \frac{f_{p}}{r_{p}} \right)
\mid {\rm \bf r}_{i} - {\rm \bf r}_{k}^{(p)} \mid
\Theta \!
\left( 
\frac{r_{p} - \mid {\rm \bf r}_{i} - {\rm \bf r}_{k}^{(p)} \mid}{\lambda} 
\right)
\hat{\rm \bf r}_{ik}^{(p)},
\label{fvp}
\end{equation}
where 
$N_{p}$
is the number of pinning sites,
$f_{p}$
is the maximum pinning force of each potential well, 
$r_{p}$
is the range of the pinning potential, 
$\Theta$ 
is the Heaviside step function, 
and 
$\hat{\rm \bf r}_{ik}^{(p)} = ( {\rm \bf r}_{i} - {\rm \bf r}_{k}^{(p)} )
/ \mid {\rm \bf r}_{i} - {\rm \bf r}_{k}^{(p)} \mid.$

The ground state of a system of moving vortices is obtained by
simulating the field-cooled experiments (e.g., \cite{tonomura-vvm}).
For deep short-range ($\delta$-like) potential wells,
the energy required to depin vortices trapped by pinning sites
is proportional to the number of pinned vortices,
$N_{v}^{(p)}$.
Therefore, in this approximation,
we can define the normalized critical current as follows 
\cite{weprl,weprb}: 
\begin{equation} 
I_{c}(\Phi) = j_{0} \frac{N_{v}^{(p)}(\Phi)}{N_{v}(\Phi)}, 
\label{jc} 
\end{equation} 
and study the dimensionless value
$J_{c} = I_{c}/j_{0}$, 
where 
$N_{v}$
is the total number of vortices within the simulation cell 
of area $A$, 
and 
$j_{0}$ is a constant 
(i.e., the current needed to depin $N_{v}$ non-interacting 
vortices, $j_{0}=N_{v}j_{v}^{dp}$).  
Since the area of the QP pattern $A_{qp}$ is less than $A$, 
the critical current Eq.~(\ref{jc}) acquires a prefactor 
$A_{qp}/A \approx 0.58$ \cite{weprl,weprb,factor} 
in the simulations below. 
We use narrow potential wells as pinning sites, 
with the maximum pinning force $f_{p}$ and the 
radius $r_{p} = 0.04\lambda$ to $0.1\lambda$.

\section{Experiment} 

Experiments were performed on $d=60\,$nm thick Nb films containing 
circular antidots (diameter $D=400$~nm) at the vertices of 
various types of QP APS (see insets of Figs.~1 and 2) 
in an area of 200$\times$200~$\mu$m$^{2}$. 
The average antidot density is $n_p\approx 0.5\,{\rm \mu m}^{-2}$, 
corresponding to a first matching field $B_1=n_p\Phi_0\approx 1\,$mT.
Electric transport measurements for the determination of the critical 
current $I_c(\Phi)$ of perforated Nb bridges were performed at 
well-controlled temperatures $T$ and magnetic fields; for details 
of sample fabrication and experimental setup see 
Refs.~[\onlinecite{koelle,kemmler-dpa}]. 
The bridges had transition temperatures $T_{c} \approx 8.4$~K and 
transition widths varying from 15 to 25~mK. 
The temperature dependence of the critical current at $B = 0$ 
differed from bridge to bridge leading to different absolute 
values of $I_{c}(B=0)$ at the same reduced temperatures $T/T_{c}$. 
The experimental data shown below, were recorded at 
$T/T_{c} = 0.9995$, 
where the absolute values of $I_{c}$ for different samples 
were very similar. 
Note that according to our estimates based on a simple core pinning 
model (see Ref.~[\onlinecite{koelle}]), the maximum value of the 
critical depinning current in quasiperiodic pinning arrays of 
antidots reached the values of about 0.5 of the Ginzburg-Landau 
depairing current, for the used parameters.

\vspace{0.2cm}

\section{Critical current in QP tiling APS} 

For a five-fold Penrose-tiling 
APS \cite{weprl,weprb,koelle,silhanek}, 
pinning arrays with many periodicities \cite{periods} 
can enhance  the critical current \cite{enhance} for a broader 
range of magnetic fluxes as compared to periodic APS. 
In order to examine and possibly further optimize $J_{c}(\Phi)$ 
in APS, in this section we study the critical current as a 
function of applied magnetic flux, for different QP tilings. 

First, we consider a tiling consisting of the simplest shapes,
squares and triangles
(both compatible with the lowest-energy vortex lattices).
The result for a quasiperiodic square-triangle tiling 
\cite{squtri} 
is shown in Fig.~1(a).
For

\begin{widetext} 

\begin{figure}[btp] 
\begin{center} 
\hspace*{-0.5cm} 
\includegraphics*[width=14.0cm]{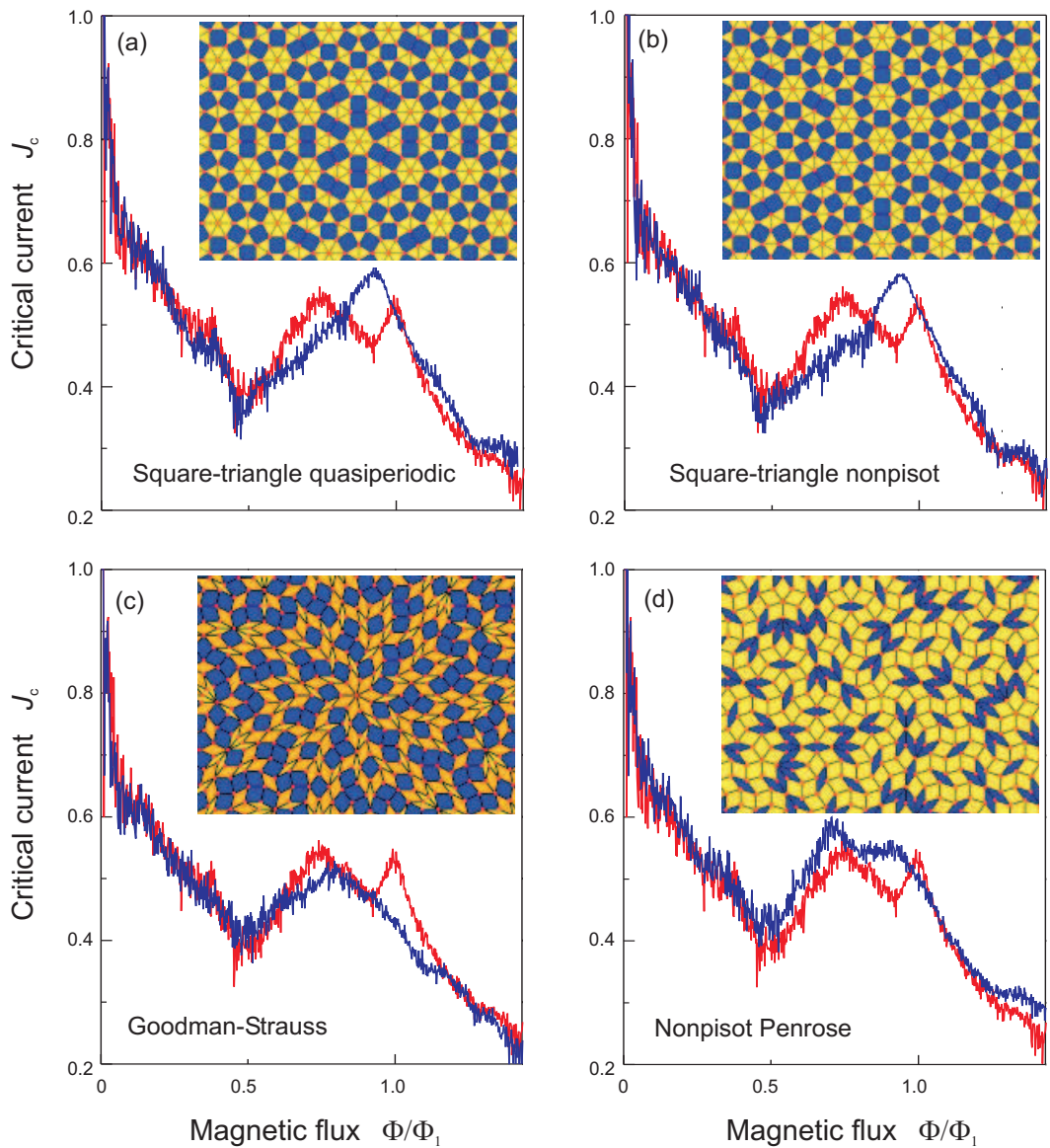} 
\end{center} 
\caption{ 
(Color online) 
Normalized 
critical current, $J_{c}$ versus 
normalized magnetic flux, $\Phi/\Phi_{1}$ 
($\propto$ number of vortices $N_{v}$), 
for different 
quasiperiodic (QP) Arrays of Pinning Sites (APS) 
[shown by solid blue (dark grey) lines]: 
(a) Square-triangle quasiperiodic tiling,
(b) Square-triangle nonpisot tiling,
(c) Goodman-Strauss tiling, and
(d) Nonpisot Penrose tiling.
Insets show the corresponding tiling patterns. 
For comparison, the function $J_{c}(\Phi/\Phi_{1})$ for a five-fold 
Penrose-tiling APS is shown by a red (grey) line. 
The $J_{c}(\Phi/\Phi_{1})$ curves are shown for $f_{p}/f_{0}=2$ 
and $r_{p}/\lambda=0.1$. 
These values of $f_{p}$ and $r_{p}$ are also used in Figs.~2-4. 
} 
\end{figure} 

\end{widetext}

\noindent 
comparison, Fig.~1 also shows the $J_{c}(\Phi)$-dependence
for a five-fold Penrose-tiling APS.
$J_{c}(\Phi)$ for the QP square-triangle tiling has a broad 
maximum at 
$\Phi_{m} < \Phi_{1}$, which is formed due to the re-arrangement 
of vortices between the square and triangular tiles when their 
density changes. 
Note the absence of a peak at $\Phi_{1}$, indicating that the 
local matching conditions between the APS and the vortex lattice 
are not fulfilled at the given parameters of the APS 
(i.e., the average distance between the pinning sites, 
their maximum strength and radius). 
However, a peak in $J_{c}(\Phi)$ at 
$\Phi_{1}$ is observed for the Penrose-tiling APS. 
Another tiling consisting of the same simple shapes 
(i.e., squares and triangles) 
is the nonpisot square-triangular tiling 
\cite{squtri} 
shown in the inset of Fig.~1(b).
Although these two types of tilings are characterizied by 
different inflation rules (i.e., different arrangements of 
the tiles), they contain the same elements, i.e., hexagons 
formed by six triangles surrounded by six squares. 
This similarity in the structure determines similar local 
matching effects between vortices and the pinning centers 
for these two tilings. 
As a result, the $J_{c}(\Phi)$-curve for the 
nonpisot square-triangular tiling [Fig.~1(b)] turns out to  
be practically identical to that for the 
quasiperiodic square-triangular tiling [compare to Fig.~1(a)]. 
An example of a tiling consisting of three different tiles
(three types of rhombuses) is a Goodman-Strauss tiling 
\cite{tilings}
shown in the inset of Fig.~1(c).
Surprisingly, the behavior of $J_{c}(\Phi)$ 
for this APS is very similar to that of the Penrose-tiling APS,
although it has an obvious shortcoming: it does not contain
the maximum at the matching flux $\Phi_{1}$ [Fig.~1(c)].

\begin{figure}[btp]
\begin{center}
\includegraphics*[width=8.0cm]{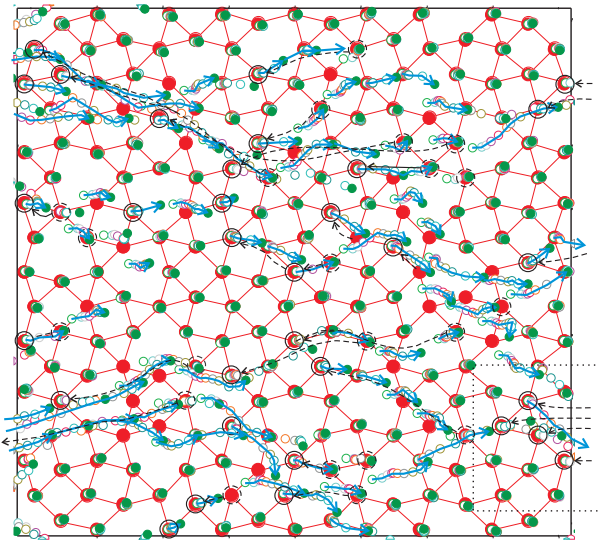} 
\end{center}
\caption{
(Color online)
Flow pattern for vortices driven by an external current 
illustrating the 
``hole-like vortex ``conductivity'' in the Shield-tiling APS 
for $\Phi \lesssim \Phi_{1}$: $\Phi \approx 0.95 \Phi_{1}$. 
The pinning sites are shown by filled large red (grey) circles. 
Vortex positions are shown for ten successive time intervals 
$t_{i}$, where ${i=1}$ to 10: 
by open small circles, for ${i=1}$ to 9, and 
by filled small green (grey) circles, for ${i=10}$. 
The Lorentz force drives vortices from the left to the right. 
Most of the vortices do not move: their positions just slightly 
change (``polarized'') with respect to the centers of the pinning 
sites. 
The trajectories of moving vortices (from the left to the right) 
are shown by thick blue (grey) arrows. 
The motion of vortices results in effective propagation of 
vacancies (``holes'') from the right to the left (shown by thin 
dashed black arrows), 
including ``holes'' entering the sample from outside, 
as in the region marked by the dotted rectangle. 
Newly created ``holes'' are shown by thin solid black circles, 
while annihilated ``holes'' (i.e., filled by vortices) are 
shown by thin dashed black circles. 
}
\end{figure}

These examples indicate that the considered tilings provide,
in general, a weaker enhancement of the critical current
as compared to the 5-fold Penrose-tiling APS which seems
to be the best among these competitors.
Let us then consider a different modification of a Penrose
tiling, namely, a nonpisot Penrose tiling
\cite{tilings}.
This tiling also consists of two types of rhombuses.
It is shown in the inset of Fig.~1(d).
As shown in the main panel of Fig.~1(d), 
this APS provides an even 
better enhancement 
of $J_{c}(\Phi)$ within nearly
the same, as for the five-fold Penrose tiling, range of magnetic 
flux: 
it is characterized by an almost-flat plateau within the 
broad maximum 
in $J_{c}(\Phi)$. 
This improvement of the pinning properties is 
explained by 
the specific arrangement of the rhombuses 
in this kind of tiling, 
which turns out to be closer to the energetically favorable 
arrangement of vortices in the lattice, thus providing better 
vortex-pin matching conditions for various densities. 
For the five-fold Penrose-tiling APS
\cite{weprl,weprb,koelle,silhanek},
the maximum
of the critical current at 
$\Phi = \Phi_{1}$ is separated
from the main broad maximum and is not as stable as the broad
maximum with respect to changes of the parameters of the APS
(i.e., the maximum pinning strength) \cite{weprl,weprb}. 
The advantage of the APS based on the nonpisot Penrose tiling
is that it provides a smooth transition from the matching-field 
configuration (i.e., at $\Phi = \Phi_{1}$) to lower-density
pinned states (i.e., when some of the vertices of the small
tiles become unoccupied \cite{weprl,weprb}).
This provides a flattening of the $J_{c}(\Phi)$-curve 
[Fig.~1(d)] and thus increases the overall value of the 
maximum $J_{c}$ within the range of the plateau-like maximum. 
This important improvement could be useful for 
applications of QP APS.

\section{The ``hole-'' and ``electron-like'' vortex ``conductivity''} 

As can be seen in Fig.~1, all the above results obtained using
different APS, while providing different profiles of $J_{c}(\Phi)$,
still have a common feature: $J_{c}$ drops drastically 
for
$\Phi > \Phi_{1}$ as it does in periodic APS (for both cases,
$\Phi > \Phi_{1}$ and $\Phi < \Phi_{1}$). 

Can we improve this?
One might think that in practice we can increase
the density of the pinning sites in the array and thus trap
flux for any $\Phi$.
Indeed this might work to some extent, but the efficiency
of an APS, especially QP APS, is determined not only by the
ratio of the average intervortex distance to the average period
of the APS ($d_{\rm av}^{\rm APS}$), 
but also by the ratio of those distances to the magnetic field
penetration depth $\lambda$, which is specific for a given
superconductor.
To understand this, let us now briefly recall
the essential difference between
vortex lattice pinning by periodic and quasiperiodic APS. 
In the former case, if the symmetry of the vortex lattice
coincides with that of the pinning array (e.g., for a triangular 
pinning array; in case of a square APS, transitions between the 
triangular, half-pinned and square lattices occur -- see, e.g., 
\cite{walterwe}) the vortex lattice and the pinning 
array are commensurate for any $d_{\rm av}^{\rm APS}/\lambda$. 
In contrast to 
periodic APS, the pinning efficiency of a QP APS 
is sensitive to this ratio $d_{\rm av}^{\rm APS}/\lambda$ 
since the elastic deformations of the 
vortex lattice are involved in its collective pinning by QP APS.

\bigskip

\begin{widetext}

\begin{figure}[btp]
\begin{center}
\hspace*{-0.5cm}
\includegraphics*[width=17.0cm]{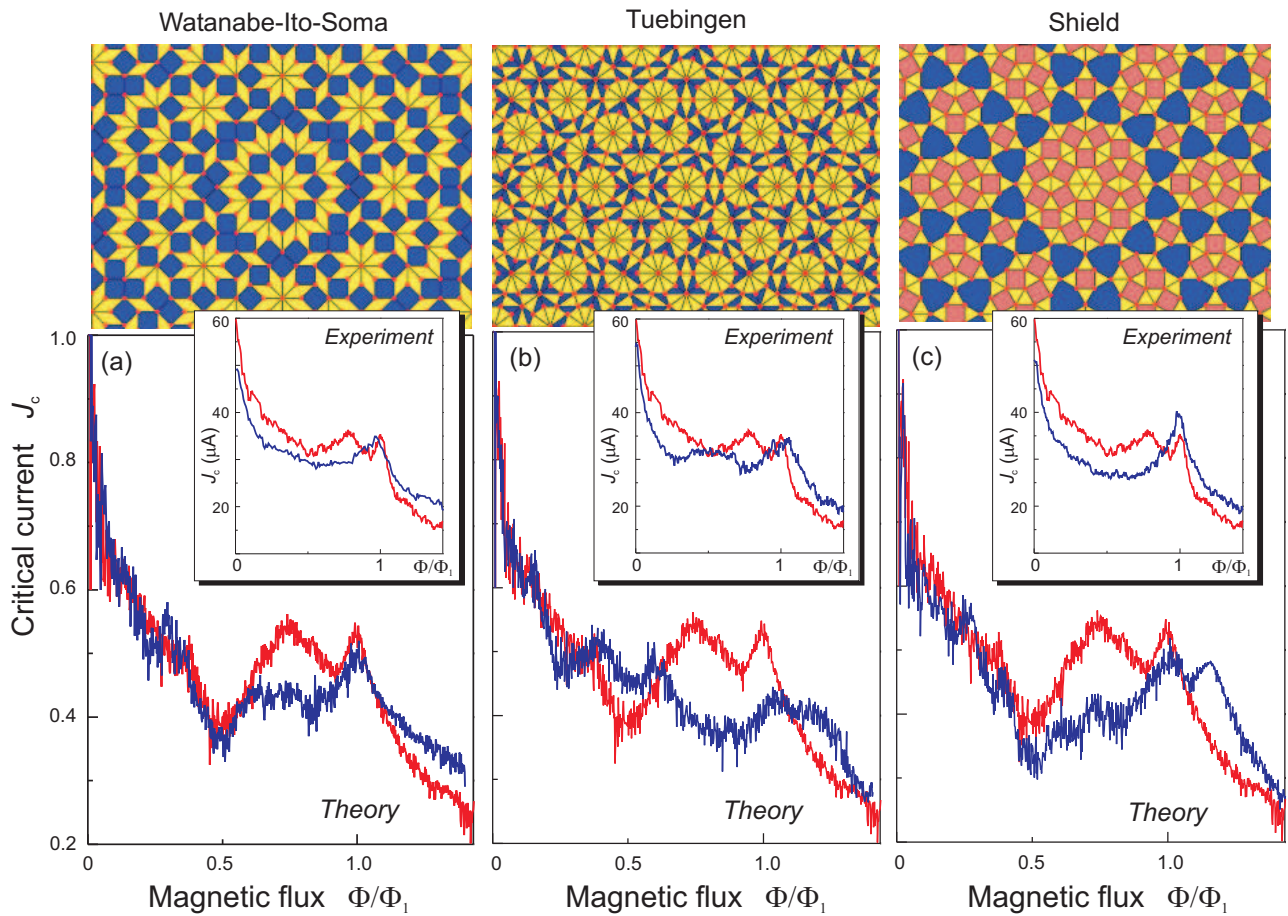} 
\end{center}
\caption{
(Color online)
Normalized critical current (experiment and theory), 
$J_{c}$ versus normalized magnetic flux, $\Phi/\Phi_{1}$ 
($\propto$ number of vortices $N_{v}$), 
for different 
quasiperiodic (QP) Arrays of Pinning Sites (APS) 
[solid blue (dark grey) line]: 
(a) Watanabe-Ito-Soma tiling, 
(b) Tuebingen tiling, and 
(c) Shield tiling. 
Insets show the corresponding tiling patterns. 
The function $J_{c}(\Phi/\Phi_{1})$ for a 5-fold Penrose-tiling 
APS is shown by red (grey) line, for comparison. 
}
\end{figure}

\end{widetext}

This is explained by the fact that a vortex lattice is
incommensurate with any QP tiling (see \cite{weprl,weprb}).
Thus, the efficiency of a QP APS does {\it not} scale with
the density of pinning sites.
On the other hand, shifting the broad maximum in $J_{c}(\Phi)$
towards larger values of $\Phi$, i.e., $\Phi > \Phi_{1}$,
results in the loss of pinning efficiency for
$\Phi < \Phi_{1}$.
Moreover, the problem of the effective pinning for magnetic
flux for $\Phi > \Phi_{1}$ is of fundamental interest and, 
as we show below, is related to a qualitatively different type 
of flux pinning in aperiodic APS. 

The fast drop of $J_{c}(\Phi)$ for $\Phi > \Phi_{1}$ is
explained by the enhanced mobility of interstitial vortices,
which is rather high even in QP APS.
Note that a very low concentration of interstitial vortices
(as compared to pinned vortices) is sufficient
to drastically reduce the critical current.
Using the language of conductivity in metals and semiconductors,
we can call the motion of interstitial vortices ``electron-like
vortex conductivity''.
Correspondingly, the situation $\Phi < \Phi_{1}$ when there are
vacancies, i.e., unoccupied pinning sites, is similar to
hole-type conductivity, since in this case the motion of vacancies
is similar to the motion of holes in semiconductors. 
For fluxes $\Phi \lesssim \Phi_{1}$, the concentration of
``holes'' is low, but the ``vortex conductivity'' is still
provided by the motion of vortices (Fig.~2).
However, in contrast to periodic arrays where this motion
is possible due to the channeling of vortices
(leading to the drop of $J_{c}(\Phi)$ for $\Phi < \Phi_{1}$),
the process of channeling is strongly suppressed in QP systems
(as in metallic quasicrystals where the usual periodic Bloch
solution for electrons does not exist 
\cite{lifshitz}).
Thus this can explain the presence of the maxima and the fast
drop of $J_{c}(\Phi)$ in QP APS, for $\Phi < \Phi_{1}$ and for
$\Phi > \Phi_{1}$, correspondingly.

\begin{figure}[btp]
\begin{center}
\includegraphics*[width=7.0cm]{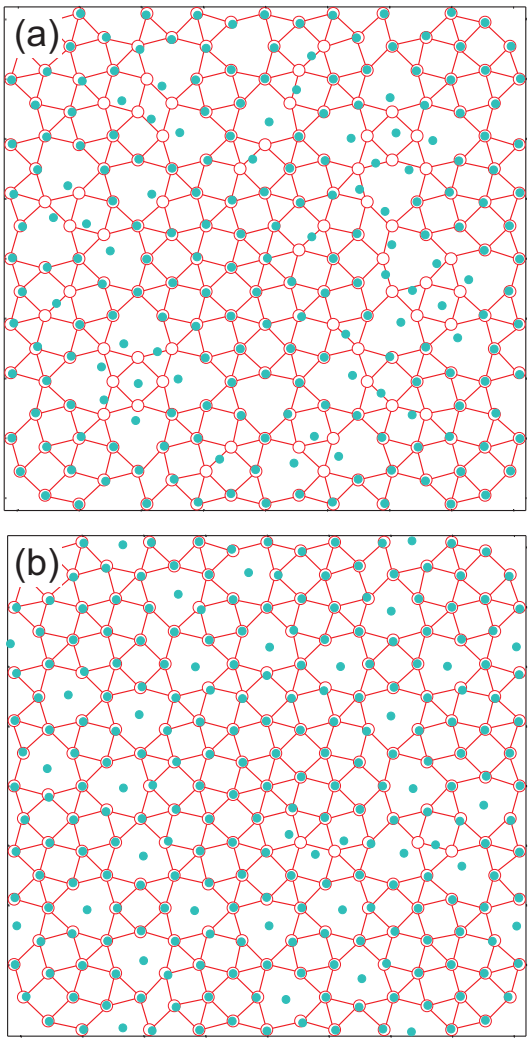} 
\end{center}
\caption{
(Color online)
Distribution of vortices in a Shield-tiling APS
calculated for:
$\Phi = \Phi_{1}$ (a) and
$\Phi = 1.18 \Phi_{1}$ (b).
The vortices are shown by green (light grey) solid dots, 
while the pinning sites by red (grey) open circles. 
}
\end{figure}

Therefore, the problem is how to immobilize the additional 
(i.e., ``electron-like'') 
vortices for $\Phi > \Phi_{1}$. 
Although in a QP APS they appear to be less mobile than in
periodic APS (no Bloch-type solutions), they can
jump to adjacent tiles (``hopping conductivity'') or ``push''
pinned vortices and thus unpin them producing secondary
interstitial vortices.
To prevent this scenario, we propose to immobilize extra
vortices in a special way, namely, to create well separated
``reservoirs'' for them which themselves are arranged
in a {\it quasiperiodic order}.
This will prevent the extra vortices from:
(i) hopping-type motion (due to the isolation of the
``reservoirs''),
and (ii) channeling (due to the QP order).

Among 2D tilings, the Watanabe-Ito-Soma tiling 
\cite{tilings,WIS}
(consisting of
squares and rhombuses) and Tuebingen tiling 
\cite{baake} 
(consisting of two types of triangles) shown in 
the insets of Fig.~3(a) and (b), 
respectively, are possible candidates.
These tilings contain isolated ``islands'' formed by triangles
and rhombuses and the density of pinning sites on these
islands is rather high.
The plots of $J_{c}(\Phi)$ in Figs.~3(a,b) show that 
indeed there is an appreciable increase of the critical current
for $\Phi > \Phi_{1}$, although the overall value of $J_{c}$ 
for $\Phi > \Phi_{1}$ for both tilings is rather low. 
These features of $J_{c}(\Phi)$ can be seen in both 
experimental and theoretical $J_{c}(\Phi)$-curves. 
Note that the experimental curves in Figs.~3(a,b,c) 
qualitatively 
(and to a certain extent quantitatively --- see Figs.~3(a,b)) 
reproduce the features under discussion of the 
calculated $J_{c}(\Phi)$-curves, 
i.e., the enhancement of $J_{c}(\Phi)$ for $\Phi > \Phi_{1}$. 
A more detailed quantitative comparison will require additional 
analysis of $J_{c}(\Phi)$ for different parameters of pinning 
arrays (e.g., radii of the antidots) and various temperatures. 
Such analysis is beyond the scope of the present work and 
will be reported elsewhere \cite{Matthias}.

A significant enhancement of $J_{c}(\Phi)$ for $\Phi > \Phi_{1}$ 
is obtained with the Shield tiling 
\cite{tilings,shield} 
(with squares, triangles, and three-fold hexagons) [Fig.~3(c)]. 
The large tiles are excellent traps for interstitial vortices 
(see Fig.~4). 
These tiles are well separated (i.e., do not have common sides) 
and are arranged in a QP order [see inset of Fig.~3(c)]. 
Thus, $J_{c}(\Phi)$ obtains a pronounced maximum of the same magnitude 
as the maximum at the first matching flux $\Phi_{1}$. 
(However the broad maximum for $\Phi < \Phi_{1}$ is suppressed, 
compared to the five-fold Penrose-tiling APS.)

\section{Conclusions}

The critical depinning current $J_{c}(\Phi)$ was analyzed
theoretically and experimentally
for different quasiperiodic-tiling arrays of pinning sites.
We showed that a five-fold Penrose-tiling APS provides
stronger enhancement of $J_{c}(\Phi)$ than other 2D QP 
tilings for a broad range of the applied magnetic flux.
It was demonstrated that the pinning properties of the
Penrose-tiling APS can be improved by using a different
modification of the tiling, i.e., nonpisot Penrose tiling.
This leads to an overall higher value of $J_{c}$ 
for a broader range of magnetic fields.
We proposed a new mechanism for the flux pinning in QP APS 
which implies an effective trap and space separation of
interstitial vortices. 
This prevents hopping-type motion and channeling of vortices 
in the APS and thus results in an increase of the critical 
current for flux values larger than the matching flux 
$\Phi_{1}$. 
These results might be useful for future applications 
in micro-electronic devices requiring a high $J_{c}$ 
over a broad range of the applied flux. 
Our proposal can be easily extended, mutatis mutandi, to other
related systems, including colloidal suspensions interacting with
pinning traps provided by arrays of optical tweezers~\cite{Grier}
or vortices in rotating Bose-Einstein condensates pinned by
potential landscapes created by co-rotating lasers~\cite{BEC}.

\section{Acknowledgments} 

This work was supported by the ``Odysseus'' Program of the Flemish 
Government and the Flemish Science Foundation (FWO-Vl), the Interuniversity Attraction Poles (IAP) Programme --- Belgian 
State --- Belgian Science Policy, the FWO-Vl, 
and by the DFG via SFB/TRR21. 
VRM is grateful to the FWO-Vl for the support of the research 
stay at the DML (ASI, RIKEN), and to Prof. Franco Nori for 
hospitality. 
MK gratefully acknowledges support from the Carl-Zeiss-Stiftung, 
and DB from the Evangelisches Studienwerk e.V.~Villigst. 
FN acknowledges partial support from the 
Laboratory of Physical Sciences, 
National Security Agency, Army Research Office, 
National Science Foundation grant No. 0726909, 
JSPS-RFBR contract No. 09-02-92114, 
Grant-in-Aid for Scientific Research (S), 
MEXT Kakenhi on Quantum Cybernetics, and 
Funding Program for Innovative R\&D on S\&T (FIRST). 



\begin{thebibliography}{0}
\bibitem{factor}
Small variation of this factor is related to slightly different 
numbers of pinning sites which fit within area $A_{qp}$ for 
differnt QP tilings.
\bibitem{periods}
Note that while QP APS contain an infinite number of periodicities, 
the corresponding diffraction patterns in the reciprocal $k$-space 
form a QP hierarchical self-similar set of $\delta$-peaks of 
decreasing intensity. 
As a result, only few periodicities turn out to be significant 
in QP APS.
\end{thebibliography}
\end{document}